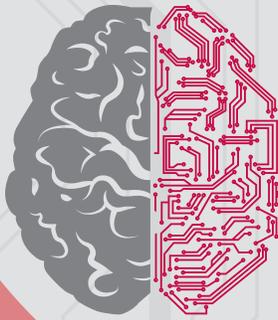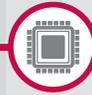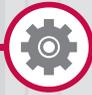

# Artificial Intelligence for Social Good

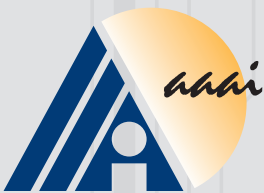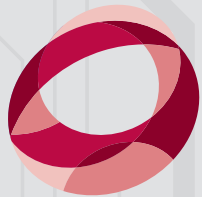

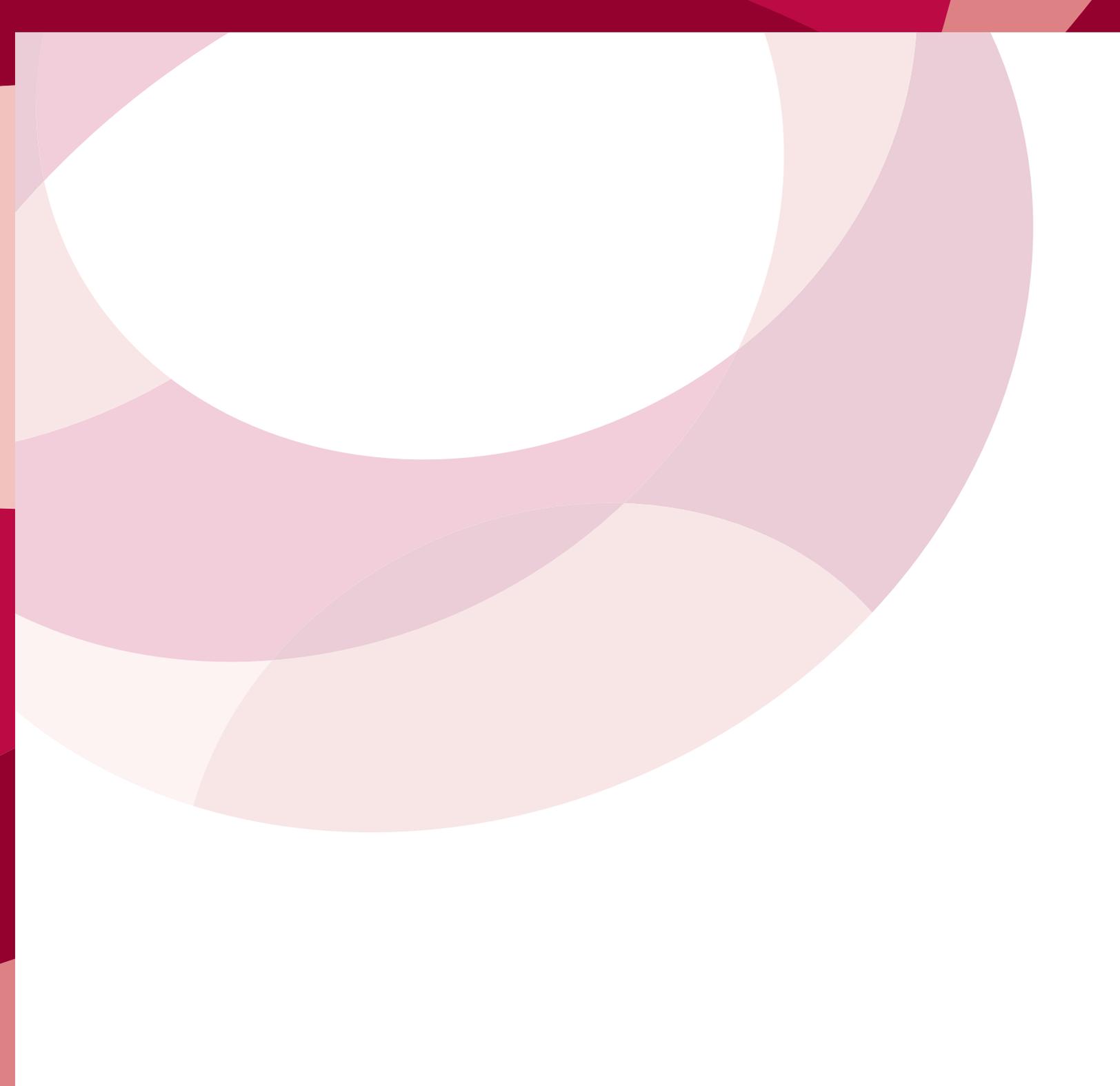

This material is based upon work supported by the National Science Foundation under Grant No. 1136993. Any opinions, findings, and conclusions or recommendations expressed in this material are those of the authors and do not necessarily reflect the views of the National Science Foundation.

# Artificial Intelligence for Social Good

Gregory D. Hager, Ann Drobnis, Fei Fang, Rayid Ghani, Amy Greenwald, Terah Lyons, David C. Parkes, Jason Schultz, Suchi Saria, Stephen F. Smith, and Milind Tambe

March 2017

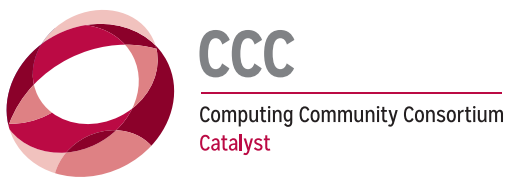
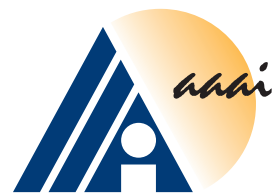

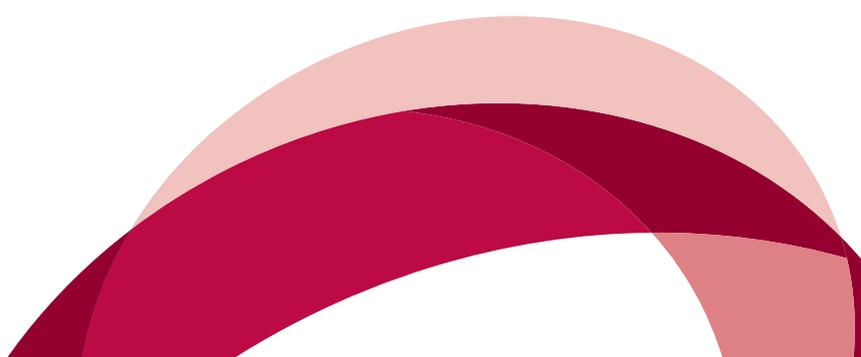

## Overview

Artificial Intelligence (AI) is currently seeing major media interest, significant interest from federal agencies, and interest from society in general. From its origins in the 1950s, to early optimistic predictions of its founders, to some recent negative views put forth by the media, AI has seen its share of ups and downs in public interest. Yet the steady progress made in the past 50-60 years in basic AI research, the availability of massive amounts of data, and vast advances in computing power have now brought us to a unique and exciting phase in AI history. It is now up to us to shape the evolution of AI research.

AI can be a major force for social good; it depends in part on how we shape this new technology and the questions we use to inspire young researchers. Currently there is a significant spotlight on the future ethical, safety, and legal concerns of future applications of AI. While understanding and grappling with these concerns, and shaping the long-term future, is a legitimate aspect of future AI research and policy making decisions, we must not ignore the societal benefits that AI is delivering and can deliver in the near future, and how our actions today can shape the future of AI.

The Computing Community Consortium (CCC), along with the White House Office of Science and Technology Policy (OSTP), and the Association for the Advancement of Artificial Intelligence (AAAI), co-sponsored a public workshop on Artificial Intelligence for Social Good on June 7th, 2016 in Washington, DC. This was one of five workshops that OSTP co-sponsored and held around the country to spur public dialogue on artificial intelligence, machine learning, and to identify challenges and opportunities related to AI. In the AI for Social Good workshop, the successful deployments and the potential use of AI in various topics that are essential for social good were discussed, including but not limited to urban computing, health, environmental sustainability, and public welfare. This report highlights each of these as well as a number of crosscutting issues.

## Urban Computing

Urban computing pertains to the study and application of computing technology in urban areas. As such, it is intimately tied to urban planning, specifically infrastructure, including transportation, communication, and distribution networks. The urban computing workshop session focused primarily on transportation networks, the goal being to use AI technology to improve mobility and safety. We envision a future in which it is significantly easier to get people to the things they need and the things they want, including, but not limited to, education, jobs, healthcare, and personal services of all kinds (supermarkets, banks, etc.).

Time spent commuting to school or to work is time not spent working, studying, or with one's family. When people do not have easy access to preventative healthcare, later costs to reverse adverse developments can far exceed those that would have been incurred had appropriate preventative measures been applied (Preventive Healthcare, 2016). Lack of easy access to supermarkets with healthful food is highly correlated with obesity (and hence heart disease, diabetes, etc.) (Studies Question the Pairing of Food Deserts and Obesity, 2012). Likewise, lack of easy access for many people to standard bank accounts is costly (Celerier, 2014). AI technology has the potential to significantly improve mobility, and hence substantially reduce these and other inefficiencies in the market to make daily living easier.

AI is now in a position to drive transformations in transportation infrastructure in urban areas. Technology exists that can mobilize people who have been immobile, due to a lack of availability of inexpensive transport; to increase flow/decrease congestion, thereby decreasing mean travel time requirements as well as variance (a great source of stress for many) (Commuting: The Stress that Doesn't Pay, 2015); and autonomous vehicles have the potential to decrease emissions (less speeding up and slowing down). The easier it becomes for people to move about, the more vibrant our urban areas will be; likewise, the more fruitful the social and economic interactions that take place inside them will be.

## Technology Enablers

The coming transformation in transportation infrastructure is being powered by technological progress. Ubiquitous connectivity and instrumentation are enabling us to measure things that were previously immeasurable; additionally, advances in data analytics are enabling us to build sophisticated models from those data. Specifically, we can now collect information about individuals' travel patterns, so that we can better understand how people move through cities, thereby improving our understanding of city life. AI technology can then be leveraged to move from descriptive models (data analytics) to predictive ones (machine learning) to prescriptive decisions (optimization, game theory, and mechanism design). Like in other domains, AI enables us to go from "data to decision" in urban computing. With the data collections now happening at this scale to aid in decision-making, it is important to also consider the privacy implications around the data.

The potential of this transformation is being demonstrated in pilot systems that optimize the flow of traffic through cities, and in new on-demand, multi-modal transportation systems. It is now within the realm of AI technology to optimize traffic lights in real time, continuously adapting their behavior based on current traffic patterns (Smith, 2016); and to dispatch fleets of small vehicles to provide on-demand transportation, address the "first and last mile" problem that plagues many urban transit systems (Van Hentenryck, 2016). More pilot deployments are needed to fully understand the scope of the transformation that is under way in our cities.

## Technical Challenges

In spite of the significant promise, many challenges lie ahead before these new opportunities can be fully realized. Transportation systems are complex, socio-technical systems that operate over multiple spatial and temporal scales. It is critical that we scale up existing pilots to multi-modal transportation models – incorporating pedestrians, bicycles, cars, vans, and buses – so that we can begin to understand how these models will impact big cities. Fundamental to this effort, it is crucial that we understand the human behavioral changes that new forms of mobility will induce, and the impact those behaviors will have on the efficacy of our systems.

## Evidence-based Policy Making

AI, as it pertains to urban computing, is in a unique position to inform policy making in ways that could not be envisioned even a few years ago. It is now possible to carry out interventions that will help us understand mobility at scale, and to analyze how different segments of the population vary their transportation modes in response to various interventions. Consequently, we are in a position to conduct research that can inform regulators, prior to the full implementation of transportation and urban planning policies. What is needed, however, is to lower the standards for testing novel AI technologies and transportation models, which may well require that we first find a way to address the psychological concerns raised by the radical transformations they promulgate.

> *Case Study: Real-Time, Adaptive Traffic Signal Control for Urban Environments.*
>
> In US cities alone it is estimated that traffic congestion costs over $160 Billion annually in lost time and fuel consumption (Schrank, 2015). Traffic congestion is also responsible for putting an additional 50 Billion tons of $CO^2$ annually into the atmosphere. A major cause of this congestion is poorly timed traffic signals. The vast majority of traffic signals run "fixed timing" plans, which are pre-programmed to optimize for average conditions observed at a particular snapshot in time and never change. These plans regularly perform sub-optimally since actual traffic flows are frequently quite different than average conditions, and they quickly become outdated over time as traffic flow patterns evolve.
>
> Recent work by Stephen Smith and his research group at Carnegie Mellon University has been applying AI techniques for online planning and scheduling to the problem of real-time traffic signal control, leading to development of the Surtrac (Scalable URban TRAffic Control) adaptive signal control system (see Fig. 1) (Smith, 2013). Surtrac senses approaching traffic and allocates green time to different approaches in real-time. It is designed specifically for optimizing traffic flows in complex urban road networks where there are multiple, competing dominant flows that shift dynamically through the day. An initial deployment of the Surtrac technology in the East end area of Pittsburgh PA has produced significant performance improvements, reducing travel times through the network by 25%, wait times by over 40%, and emissions by 21% (Smith, 2013). Over the past 3 years, this Pittsburgh deployment has grown to an interconnected network of 50 intersections, and the City of Pittsburgh currently has plans and funds in place to further expand and equip an additional 150 intersections with this technology.

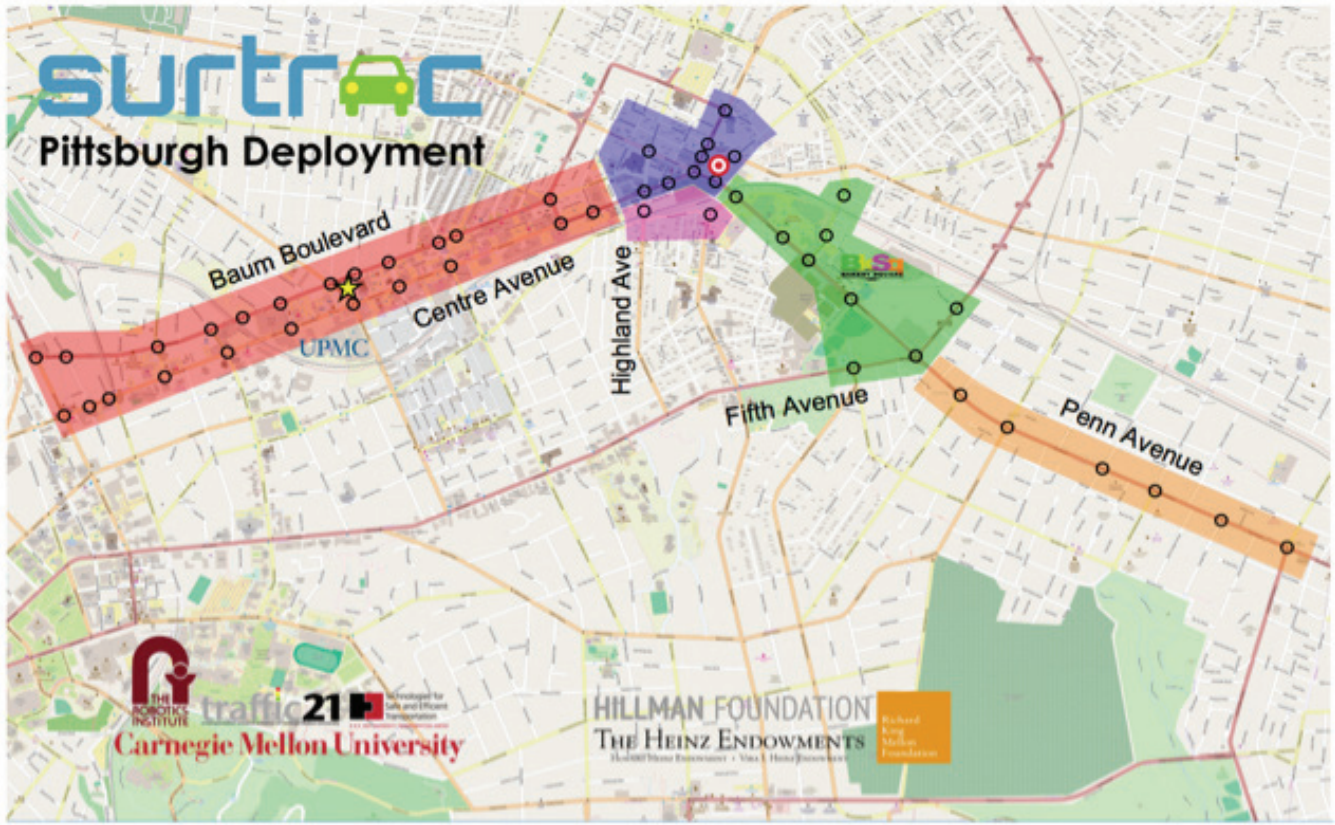

*Figure 1: Scalable URban TRAffic Control*

Current research with Surtrac focuses on integration of smart signal control with emerging Dedicated Short Range Communication (DSRC) radio technology (Smith, 2016). This "connected vehicle" technology, which will begin to appear in some makes of new passenger vehicles in the US starting in the 2017 model year, will allow direct "vehicle-to-infrastructure" (V2I) communication. In addition to simple use of V2I communication to promote safer travel (e.g., through advance warning of pending signal changes), projects aimed at utilizing V2I communication to enhance urban mobility (particularly under the shorter-term assumption that the penetration level of equipped vehicles is low) are also underway.

## Sustainability

Sustainability can be interpreted narrowly as the conservation of endangered species and the sustainable management of ecosystems. It can also be interpreted broadly to include all aspects of sustainable biological, economic, and social systems that support human well being. Here we focus primarily on the ecological component, but the larger issues of social and economic sustainability must be considered as well.

**Short-term Applications and Challenges**

Current research and applications in AI for sustainability can be organized in terms of data, modeling, decision making, and monitoring. The goal is to manage ecosystems with policies that are based on the high quality data and science.

**Data**

Several activities concern the measurement and collection of data relevant to ecosystems.

One approach is to develop and deploy sensor networks. For example, the TAHMO (Trans-Africa Hyrdo-Meteorological Observatory; www.tahmo.org) project is designing and deploying a network of 20,000 weather stations throughout sub-Saharan Africa (van de Giesen, 2014). Several efforts are deploying camera traps to collect image data or microphone systems to collect bioacoustic data. Still other projects employ unmanned aerial vehicles to obtain video imagery for tracking elephants and other large animals. AI algorithms can be applied to optimize the locations of these sensors and traps in order to gather the most valuable information at the lowest cost. Once the data are collected, other AI algorithms can be applied to identify species and track their locations.

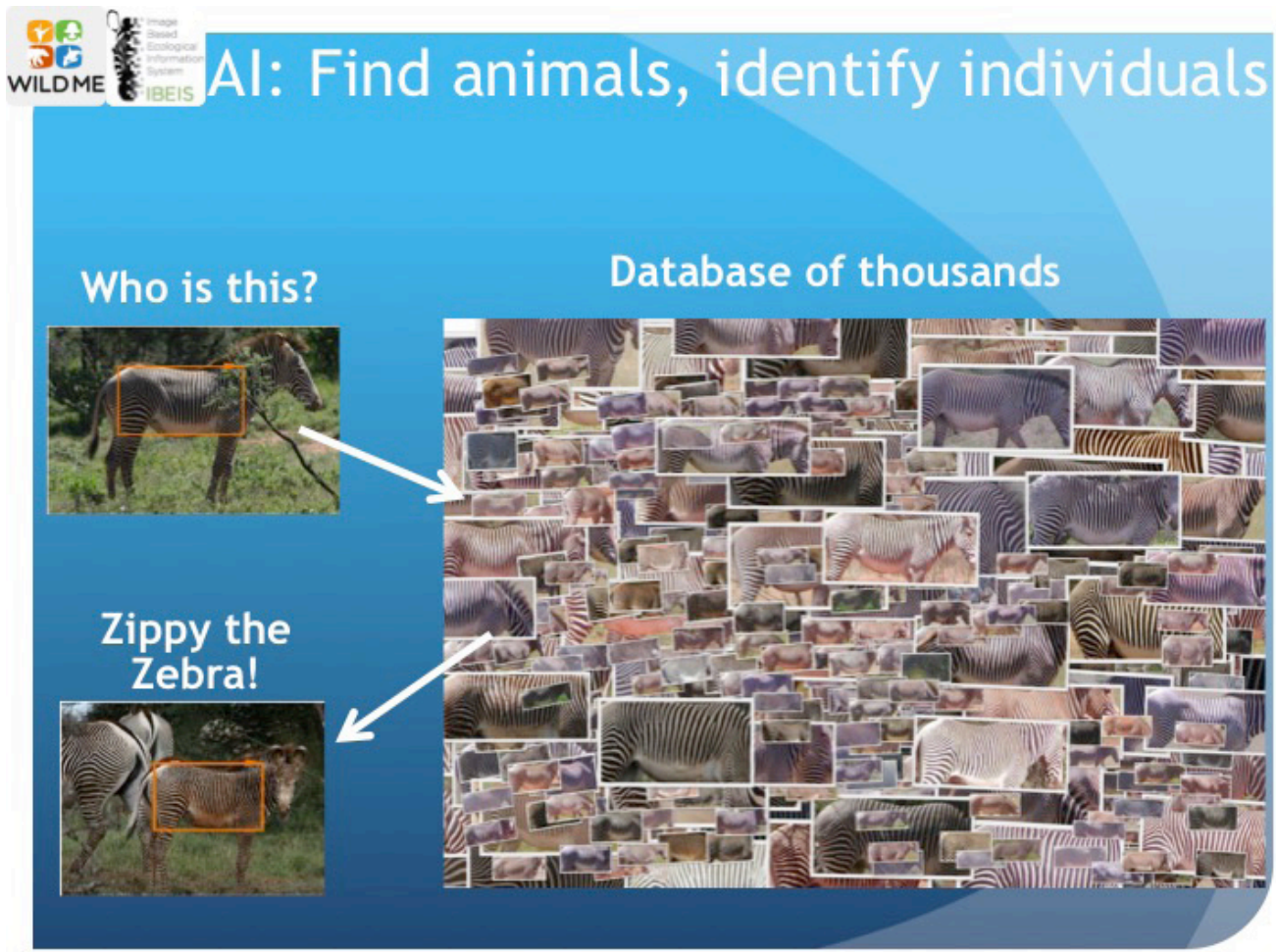

*Figure 2: Image-Based Ecological Information System*

A second approach to data collection is to engage citizen volunteers. One of the oldest citizen science projects is eBird (www.ebird.org), in which bird watchers upload checklists of the birds they have seen at a particular time and place. The Image-Based Ecological Information System (IBEIS; www.ibeis.org) project analyzes animal photos scraped from internet sources such as Flickr and Facebook and applies computer vision and active learning methods to detect the animals, identify the species, and even identify individual animals ("Bob the giraffe") (see Fig. 2). Their AI techniques can identify unique animals as long as they have stripes, wrinkles, or other unique textures.

A third approach employs technically trained people (e.g., government and corporate scientists) to collect data. One example of this is the freshwater stream surveys conducted by the EMAP project of the EPA (https://archive.epa.gov/emap/archive-emap/web/html/), and similar efforts by forest resource companies such as Weyerhauser (http://www.weyerhaeuser.com/timberlands/forestry/about-our-forests/us-west/). These groups collect samples of freshwater macroinvertebrates that live in streams. These insects must then be examined by humans to identify the genus and species of each. AI computer vision methods are now being applied to accelerate this process.

## Models

After data are collected, the data can be analyzed by applying techniques from data mining, statistics, and machine learning to discover trends and fit models. For species data, most efforts begin by counting individuals in order to produce estimates of population size and maps of the spatial distribution of species. For this purpose, it is particularly valuable to identify individuals and take into account multiple detections of the same individual across time and space. These models can support some inferences about species habitat requirements. In light of climate change, an important goal is to understand which climate variables affect species habitat carrying capacity.

A second important type of model seeks to characterize the spatio-temporal dynamics of species. Such models can predict migration, dispersal, reproduction, and mortality of species. Developing such dynamical models is critical to developing policies that can help endangered species thrive and control the spread of invasive species. One recent example is the BirdCast bird migration model under that combines eBird and weather radar to test hypotheses about the behavior of migrating birds (birdcast.info).

## Policy Optimization

Once we have models of species distribution, behavior, and habitat requirements, we can begin to design and optimize policies for successful management of species and ecosystems. This requires articulating our policy goals and objectives. Virtually every ecosystem management problem combines an ecological model with an economic model of the economic costs and benefits of various policy outcomes.

One example is the design of a schedule for purchasing habitat parcels to support the spatial expansion of the Red Cockaded Woodpecker (Sheldon, 2010; Sheldon, 2015). An optimal policy takes into account uncertainty in the spread of the species and the availability of land parcels while seeking to link up existing patches of reserved habitat. Algorithms for computing this policy combine ideas from network cascade analysis (maximizing spread in social networks) with techniques from AI planning and Monte Carlo optimization.

A second example considers the temporary needs of migrating birds. Instead of making a permanent purchase of land, the Nature Conservancy is applying detailed bird migration models developed by the Cornell Lab of Ornithology to rent rice fields in California (Axelson, 2014). The farmers who own those fields agree to flood them at the right time to support migrating waterfowl in the pacific flyway. The exact timing varies from year to year based on the predictive migration models developed using multi-scale machine learning techniques (cite STEM).

A third example of policy development confronts the issue of long-term planning in the face of climate change and sea level rise. One approach, known as adaptive management, explicitly considers the need to update and revise models based on new data that will become available in the future. There is considerable uncertainty about the timing and degree of sea level rise. Nicol, et al., (2015) formalize the problem of planning in the presence of this uncertainty. Their goal is to conserve appropriate coastal habitat for migrating birds under the risk of sea level rise. Low-lying land will become inundated, and migrating birds will be threatened unless additional habitat further inland is available for them. As always, there are tight budgetary constraints on the amount of land that can be purchased. AI algorithms for solving Partially-Observable Markov Decision Problems are able to solve these difficult constrained optimization problems (Pineau, 2003).

> *Case Study: Monitoring and Enforcement*
>
> When policies are put into effect, there is often a need for law enforcement to ensure their successful execution and for monitoring to detect errors in the data and models that require re-optimizing the policy. For example, elephant poachers in Africa routinely enter national parks and other bioreserves to hunt elephants illegally for the ivory tusks. The PAWS project (Fang, 2016; Nguyen, 2016; Yang, 2014) applies AI algorithms to predict poaching attacks and optimize the patrol routes of game wardens in order to maximize their deterrent effect while minimizing costs. PAWS relies on two key underlying systems: (i) *Predicting poacher behavior from past poaching data:* This system builds poacher behavior models using machine learning algorithms (Nguyen, 2016); (ii) *Game theoretic security resource allocation:* Using these learned poacher behavior models, it uses game theoretic algorithms to prescribe new patrolling strategies; thus, rather than assuming the standard rational adversary response in standard game theoretic models, this work assumes that the adversary behavior is governed by models derived from past adversary behavior. The use of security resource allocation using game theory has been explored in urban contexts, particularly in counter-terrorism settings in the past (Tambe, 2011). The current work builds on this past work while also significantly enhancing it with machine learning based predictions of adversary behaviors.

## Opportunities and Challenges in the Medium Term

With few exceptions, most work in ecosystem management and conservation focuses on a small number of species in particular regions. A major challenge for the medium term is to develop methods that can collect and model data encompassing a broad range of species at continental scales. This will require integrating many different data sources (e.g., for birds, fish, plants, and insects) collected by many different methods (e.g., stationary sensors, earth orbit satellites, citizen scientists, sensors worn by animals, and so on). There are many research issues in data management and data integration that must be addressed. For example, the most common approach to data integration is to assimilate all data to a fixed spatial and temporal scale by smoothing fine-scale data and interpolating coarse-scale data. This process introduces distortions into the data. We need methods for integrating and modeling data at multiple scales that can retain the resolution and uncertainty associated with each data source.

A related shortcoming of current modeling efforts is that they generally assume stationary (steady-state) climate, land use, and species behavior whereas the real systems are experiencing climate change, rapid economic development, and continuing evolution, dispersal, and natural selection of species. Modeling techniques and supporting data are needed that can take into account these drivers of change and the many uncertainties associated with them.

As the scale of questions grow, it is no longer possible to focus only on the biological components of a system. Instead, one must take a "systems of systems approach" and incorporate models of social, cultural, and economic activity. For example, when choosing a site for a new dam, we must consider not only the impact on native and invasive species in the riverine ecosystem, but also the benefits for farming, the potential inundation of important cultural and religious sites, and changes in sediment transport that may affect the distribution of pollutants and contaminated soils. Current AI technologies cannot currently operate at this scale and level of complexity.

One trend that will enable broader and more comprehensive modeling of ecosystems is the continuing improvement of sensors: reduction in size, power requirements, and cost. These improvements will support and drive the demand for better models that can support the development of higher-quality policies. However, cheaper sensors can be less reliable, so research is needed on methods for automatically detecting and removing bad data and broken sensors. This is a theme that is also common to other areas such as healthcare and public policy, as we discuss below, and poses a major challenge for many AI-powered decision-support systems.

A second set of data challenges concerns the biases and quality of data, particularly crowd-sourced data. Birders choose where they go bird watching; tourists and tour operators choose where people take pictures of wildlife. Even the data collected by game wardens is biased by the need to maintain unpredictability. New incentive mechanisms are needed to encourage volunteers to collect less biased data. Examples of mechanisms that are showing some success include the Great Zebra Challenge and the eBird Global Big Day. New algorithms are needed to incorporate data collection goals into the PAWS enforcement games. And methods for explicitly modeling the data collection process ("measurement models") must be improved. A major analytical challenge is that when measurement models are incorporated into machine learning, the variables of fundamental interest are no longer directly observed. This raises questions about the identifiability and semantics of the inferred values of those variables.

A third challenging aspect of sustainability work arises due to lack of technical infrastructure: poor networking, little access to high-performance computing resources, and lack of local personnel with sufficient education and training. We must develop algorithms that can run locally on small computers (or telephones) that only have intermittent access to large cloud computing resources. We must take into account the possibility that human actors may fail to adhere to designated policies. Finally, we must develop creative methods of establishing metrics for assessing the effectiveness of data collection and policy execution to compensate for the lack of historical data.

A fourth challenge is finding business models that support long-term data collection and policy enforcement efforts. Many current projects rely on the enthusiasm of citizen scientists, the generosity of private donors, or grants from funding agencies. None of these is likely to provide steady, long-term support. One possibility is to develop business models that generate continuing revenue streams. For example, the TAHMO project seeks to sell its weather data to insurance companies, commodities traders, and other businesses that rely on high quality weather data and forecasts.

## Long Term Prospects

Sustainability is concerned with the long-term health of ecosystems and human societies. As we contemplate the creation and deployment of policies over the long term, we must confront the fact that the long-term behavior of ecological, economic, and social systems is radically uncertain. We can be very confident that our current models are missing critical variables and important interactions. How can artificial intelligence methods deal with the uncertainty of these "unknown unknowns"?

One important strategy is to plan for the "learning process". When a new policy is put into place, we must also develop and deploy an instrumentation plan to collect data on a broad range of variables. We must incorporate "precautionary monitoring", in which we monitor not only the variables that we expect to change as a result of the policy, but also a wide range of variables that could allow us to detect unexpected side effects and unmodeled phenomena. We must plan to iteratively extend our models to incorporate these phenomena and re-optimize the policies.

Finally, when formulating and optimizing management policies, we should adopt risk-sensitive methods. Standard practice in solving economic models is to minimize the expected costs and maximize the expected benefits of the policy. But if a policy has substantial downside risk (e.g., species extinction, economic catastrophe), then we should apply AI methods that find robust policies that control these downside risks. This is an active area of research (see, e.g., Chow, 2015), and much more work is needed to understand how we can ensure that our models are robust to both the known unknowns (as in traditional risk management methods) and the unknown unknowns.

# Health

**Success Stories**

Current methods for gathering population-scale data about public health through surveys of medical providers or the public are expensive, time consuming, and biased towards patients who are already engaged in the medical system. *Social media analytics* is emerging as an alternative or complementary approach for instantly measuring the nation's health at large scale and with little or no cost. Natural language processing can accurately identify social media posts that are self-reports of disease systems, even for rare conditions. The nEmesis system, for example, helps health departments identify restaurants that are the source of food-borne illness (Sadilek, 2016). nEmesis finds all the Twitter posts for a city that are sent by restaurant patrons, and then checking if any of the patrons tweet about the symptoms of foodborne illness over the next 72 hours. When this happens, health department officials are alerted of the fact, so that they can schedule inspection of the restaurant. nEmesis significantly improved the effectiveness of inspections in Las Vegas and the Center for Disease Control is funding the expansion of nEmesis nationwide.

The *Surgical Critical Care Initiative* (SC2i), a Department of Defense funded research program, has deployed two clinical decision support tools (CDSTs) to realize the promise of precision medicine for critical care (Belard, 2016). The *invasive fungal infection* CDST was deployed in 2014 to assist military providers with treatment decisions both near point of injury and at definitive treatment centers. Trauma-related invasive fungal infections are well recognized for their devastating impacts on patients in both military and civilian populations. In addition to substantial morbidity resulting from recurrent wound necrosis (e.g., greater number of surgical procedures, amputations, and delayed wound closure), the disease is also associated with high mortality rates (Tribble and Rodriguez, 2014; Warkentien, 2012; Lewandowski, 2016; Rodriguez, 2014).

The *massive-transfusion protocol* (MTP) CDST is currently being assessed under a two-year clinical trial at Emory-Grady, one of the two SC2i civilian hospitals. This CDST uses evidence-based predictive analytics to help physicians identify which patients genuinely require a massive transfusion, thereby reducing complications associated with over-transfusion or the needless expenditure of blood products (Maciel, 2015; McDaniel, 2014; McDaniel, 2014; O'Keeffe, 2008; Dente, 2009). The SC2i is also planning a clinical trial around its *WounDX*, a CDST that predicts the timing of traumatic wound closure. Once validated, this tool has the potential to substantially improve outcomes (by as much as 68%) and reduce resource utilization ($3.4B annual cost-savings) both nationally and within the Military Health System (Forsberg, 2015).

> *Case Study: Making 'Meaningful Use' meaningful*
>
> Sepsis is the 11th leading cause of death in the US – seven hundred fifty thousand patients develop severe sepsis and septic shock in the United States each year. More than half of them are admitted to an intensive care unit (ICU), accounting for 10% of all ICU admissions, and 20% to 30% of hospital deaths. Yet others experience sepsis due to hospital acquired infections (HAIs) in the medical units. Several studies have demonstrated that morbidity, mortality, and length of stay are decreased when severe sepsis and septic shock are identified and treated early; Kumar et al. 2006 show that every hour delay in treatment is associated with a 7-8% increase in mortality. Screening tools exist but these typically implement guidelines that are based on clearly visible symptoms, often delaying detection.

Alternatively, the use of biomarkers or specific lab tests delay detection until caregivers are suspicious of decline. Recent work has led to new automated real-time surveillance tools: by using analytic techniques that integrate diverse data – routinely collected in the electronic health record – these tools identify individuals at risk for severe sepsis and septic shock at the early stages of decline, and much earlier than standard of care (Henry, 2015). Early warning opens up the possibility for providing the sepsis bundle in a timely fashion, which has been shown to reduce mortality rates by more than 50% (Barwell, 2014). Similar ideas have been explored for risk monitoring of individuals likely to test positive for C-diff (Wiens, 2014). In yet another example, in neonatology, routinely collected physiological data streams have been shown to construct an electronic score to risk stratify premature newborns into low-risk and high-risk cohorts (Saria, 2010).

## Near Term Opportunity

### 1) Targeted therapy decisions

Many chronic diseases are difficult to treat because of high variation among affected individuals. This makes it difficult to choose the optimal therapy for a patient. Developing systems that support targeted therapy decisions from large-scale observational data is an emerging and exciting area of research (Murphy, 2003). By analyzing longitudinal databases of clinical measurements and health records, we can develop *decision support tools* to improve decision-making. This can take us towards *precision medicine*.

*Computational subtyping*, for example, seeks to refine disease definition by identifying groups of individuals that manifest a disease similarly (Collins, 2015). These subtypes can be used within a probabilistic framework to obtain individualized estimates of a patient's future disease course. Better decision support tools can be used for more than improved disease management – also providing for better wellness and diagnosis.

### 2) New sensors, new healthcare delivery

AI can be used to analyze *social media data* and discover and suggest behavioral and environmental impacts on health. In addition to the *nEmesis* system described above, examples include tracking influenza and predicting the likelihood that particular social media users will become ill, and quantifying alcohol and drug abuse in communities. Social media as well as *social networks* can also be used to address the informational and psychosocial needs of individuals e.g., the American Cancer Society's *Cancer Survivor Network* (CSN) (Bui, 2016). Related also is the opportunity for cost-effective interventions for addressing mental health, addiction, and behavioral health issues using modern low cost sensing technologies. Data gathered routinely during healthcare delivery can be leveraged to reduce hospital-acquired infections. *Low fidelity sensors*, some of which are diagnostic, together with AI and internet technologies can enable *low barrier telemedicine* for example for chronic healthcare. Advances in natural language processing and machine reading can be used to synthesize, integrate and appropriately disseminate new medical knowledge (e.g., as reported in journal articles.)

## Near Term Enablers

**Incentive Alignment:** We should consider questions of incentive alignment in order to encourage various actors in the health ecosystem to collect additional data and make their data available to the rest of the healthcare ecosystem (this includes health care providers as well as payers such as health insurance companies). For example, providing hospitals that share data with an immediate benefit such as predictions that improve their use of hospital resources, thus enabling operational improvements. We should also provide clarity on the additional data that would be useful for health providers to collect, in order to provide better predictability for the effectiveness of therapy decisions. At the same time, we should address biases in public health data, for example countering biases by modeling the data acquisition process, and by encouraging the self-reporting of additional data.

**Data Science Platforms:** Cloud-based data sharing and common data models should be developed and promoted in order to increase the likelihood of societally beneficial outcomes. Shared experimental test beds will also be important, in order to lower the barrier of entry into AI and health research in order to bring in more researchers. Another direction is to utilize non-U.S. data to speed up the development and testing of models.

## Longer Term Opportunity

**Personalized Health:** In this time frame, the major opportunity is to pivot from personalized medicine to personalized health, to keeping people from getting to the hospital in the first place, and to dealing with life issues and not just specific diseases. For this, we need to move to modeling the health of individuals and populations by using integrated data sets – electronic health records data and other data gathered within the health system with genomic, socio-economic, demographic, environmental, social network and social media and other, non-traditional data sources, such as social service and law enforcement data. Particularly relevant in this context are causal inference methods (Pearl, 2000), including methods for inferring causal effects from disparate experimental and observational studies (Bareinboim, 2013; Bareinboim, 2014; Lee and Honavar, 2013a; 2013b; Pearl, 2015) and from relational data (Maier, 2010; 2013; Lee and Honavar, 2016a; 2016b; Marazopoulou, 2015). From this can come personalized, longitudinal treatment plans to improve an individual's health.

**Collaborative Decision-Making:** We need approaches that allow decision makers to collaboratively reason with models of the health of individuals. For example, can a healthcare provider ask a question about how the trajectory of an individual's disease would change if a test came out positive? How would this health trajectory change if the individual was being treated with two different drugs? These questions can help a decision maker develop a mental model of the computational system and learn to use its output to influence decisions. What are good frameworks that integrate messy data as it arrives, maintain estimates of uncertainty, and support flexible, collaborative decision-making?

**Addressing Bias:** An important challenge that arises in fitting models from observational health data sources is that the data may be influenced by many phenomena, including some unrelated to the target disease of interest and arising from the structure of the health ecosystem. For example, information may be missing on conditions that are not reimbursed. Not accounting for these biases can lead to models that cause harm as provider practice patterns change (DS2016, PNS203). We also need to measure the extent to which models are transportable: building tools from health data requires models that are "healthcare process aware" (HA2013, BLHP2013, SW S2015) – in other words, what is the measurement process and how did it affect the data that were generated?

## Barriers to Progress

**Need for Cross-Disciplinary Training:** We need programs that train scientists in developing AI methods for complex socio-technical systems. Beyond training in methods from computer science and statistics, these individuals must gain exposure to working with domain scientists to understand problem requirements. For example, we need new frameworks for measuring performance. Contrary to existing ways where emphasis is on measuring performance with fixed datasets, our metrics must measure accountability and reliability in "living systems" – environments that are constantly changing. Scientists must also be trained in the ethics of working with human subjects data.

**Privacy:** We need better methods to understand how to work with data in a way that both sustains its utility while the decisions and outcomes of working with the data do not reveal information about individuals that can lead to a loss of privacy. A particular challenge is that the validity of methods of perturbation or aggregation, both in regards to privacy and retaining the utility of data, depends on task context. That is, protecting privacy while enabling utility is an important challenge. We need policy portability so that data can move around but still be used correctly, and connected with this is the need for methods to audit data use and to reconcile policy across organizations. Related to this, is the need for policy in regard to the data custody (ownership and management) of electronic medical records, including rules for access and commensurability (risk vs. value vs. penalty). Security is also vitally important and presents a barrier unless technological solutions can be put in place – personalized medicine requires an integrated view of an individual and this data must be kept secure.

### Grand Challenge

The grand challenge for AI and health is to develop a learning healthcare system. This is a sustainable system that is able to observe all the available data about a person, build appropriate models from the observations, help make the right decisions using all available AI technologies, proactively and reactively make the right interventions and care when the person needs it, re-capture the results of the intervention, and learn and adapt from the feedback.

## Public Welfare

Recent advances in Artificial intelligence have resulted in impact on several industries such as retail, security, defense, manufacturing, transportation, search, social networking and advertising. At the same time, AI has not had a lot of impact on fundamental issues our society faces today. Education, public health, economic development, criminal justice reform, public safety are just some of the areas where AI can potentially make an impact. Here we will discuss some of the issues in these areas where AI can play a critical role, describe some early work where AI has already started to make an impact, and highlight fundamental issues, short and long-term opportunities, barriers and grand challenges for AI research applied to public good. The overall end goal of the work we're motivating is to enhance the quality of life for all individuals, and increase equity, effectiveness and efficiency of public services being provided to citizens.

### Social Issues AI can help address

There are many social issues AI could contribute to. Some examples include the following:

**Justice**: How do we identify, target, and prevent individuals who are likely to cycle through various public systems (emergency rooms, homeless shelters for example) and eventually end up in the criminal justice system? Can I understand what factors best predict interactions with these systems and develop interventions so employees of these systems may reduce future interactions while providing quality services?

**Economic Development:** How do I allocate resources that a city has towards the homes, neighborhoods, and communities that are most likely to help reduce blight?

**Workforce Development:** How do we help Job Training and Skills Development programs to figure out what skills are going to be in demand in the future so they can train individuals and help them become employable?

**Public Safety:** How do I better make dispatch decisions for emergency response calls? Who do we send for a given dispatch and how do we ensure to send the appropriate resources without overspending?

**Policing:** Can I identify police officers that are at risk of adverse incidents with the public in order to match them appropriate preventative interventions?

**Education:** Build a system to help target early and effective interventions at students who may need extra support to graduate on time or not likely to apply to college or not ready for college or careers upon graduation from high school

## Success Stories and Case Studies

There has been recent work on some of the issues described above, mostly done by universities in collaboration with government agencies. Some examples include the following:

**Public Health:** Targeted Lead Inspections to Prevent Child Lead Poisoning: Almost 9,000 children in Chicago are thought to have been exposed in 2013 to lead at levels classified as dangerous by the CDC. While most of this exposure happens in the child's home, limited funding and personnel make it impossible to inspect all 200,000 Chicago buildings built before laws banned lead paint. This is the case in many cities, where lead remediation only occurs after a child in the home presents dangerous blood lead levels. The University of Chicago partnered with the Chicago Department of Public Health to build a system (Potash, 2015) to predict which children are at risk of lead poisoning, before they get exposed to lead, and allow CDPH to deploy inspectors and proactively address lead hazards before exposure impacted more children. CDPH is currently running a trial to validate the performance of the model as well as implementing it into the inspection targeting system and the Electronic Medical Records systems of hospitals so these risks can be determined while a woman is pregnant and remediation can happen before the child is born.

**Education:** Helping students graduate on time: Over 700,000 students drop out of high school every year in the United States. High school graduation is associated with relatively higher overall lifetime earnings and life expectancy, and lower rates of unemployment and incarceration. Interventions can help those falling behind in their educational goals, but given limited resources, such programs must focus on the right students at the right time and with the right message. Over the past several years, several school districts around the US have been collaborating with universities to develop AI based systems to help them identify at-risk students who are unlikely to finish high school on time (Lakkaraju, 2015).

**Public Safety:** Recent high profile cases of law enforcement officers using deadly force against civilians and other instances of police misconduct have caused a growing political and public uproar. The University of Chicago Center for Data Science and Public Policy has been working, as part of the Obama White House Police Data Initiative, to address these problems using a data-driven and predictive approach (Carton, 2016) - to identify officers who are at risk of adverse incidents early and accurately so supervisors can effectively target interventions. This system takes data about officer demographics, training, payroll, on-the-job actions, internal affairs data (complaints, investigations, reviews of incidents, etc.), dispatch data, negative interaction reports as well as some publically available data and uses machine learning methods to assign each officer a risk score. Compared to the existing Early Intervention System, the AI based system can correctly identify 10-20% more officers who go on to have adverse incidents over a 12-month period while reducing the false positives by 50-60%. This improved capability allows police departments to effectively and efficiently identify at-risk officers and provide them the necessary preventative interventions before an adverse incident occurs. Early interventions can lead to fewer adverse police interactions with the public, reduce injuries sustained by citizens and officers, improve the wellbeing of officers, improve police community relations and overall policing across the US.

> **Case Study: Economic Development:** Targeted Home Inspections to Reduce Urban Blight
>
> Blight starts as a small problem, usually only one house in a neighborhood starts to run down. However, it spreads fast, up and down the street and through other neighborhoods. First neighborhoods start declining, then communities start declining as people become unemployed and move away. Usually cities wait until an entire neighborhood has been affected, then they start to look for investments that they can put back into the neighborhood to try and revitalize them. Sometimes it works, but most of the time it is just too expensive to bring them back.

> Rayid Ghani from the University of Chicago Center for Data Science and Public Policy and his researchers have been working with the City of Cincinnati to combat this problem. They are taking the preventative route and looking at the past 10 years of data on which areas are prone to blight and starting to predict which neighborhoods and homes will be next. Targeted home inspectors can then be sent to houses before blight happens.

## Gaps and Barriers

Work in this area requires deep, intimate, and sustained interaction and efforts between the target community and AI researchers. Some of the current gaps and barriers are as follows:

**Lack of Experienced Collaborators:** There isn't an established history of AI working in this area. As a result, there isn't a ready supply of trained AI researchers (or practitioners) who are familiar with the unique aspects of working on public welfare problems. Conversely, government and policymakers have little experience working directly with the research community. Highlighting ongoing projects (and successes) to both raise awareness and to provide a roadmap is essential to growing this community. Also critical is training both sides on how to scope and formulate problems and projects that result in effective collaborations and impact. Several training programs targeted at seeding and fostering these collaborations have been created over the past several years, such as the Data Science for Social Good Fellowship program, initiated at the University of Chicago in 2013, and replicated by Georgia Tech, University of Washington, and IBM.

**Lack of Visible Activity and Case Studies:** Building on the previous point, the level of activity in this space is far lower than the needs. This lack of activity makes it difficult for governments and policymakers to know what's possible when thinking about the uses of AI in their work. Increasing projects in this area is a "retail" problem, as pilot projects will inevitably be local and therefore shaped by the unique context and capabilities of the municipality and research group. Finding funding mechanisms that address local needs – e.g. the NSF Data Hubs model – is essential. Seeding lots of small prototype projects is also critical at this stage.

**Lack of Reusable Infrastructure:** While AI tools are increasingly available to a broad set of researchers, the underlying platforms to support them, within a context of public good, is missing. To continue the previous example, identifying at-risk populations will require access to data sets such as tax records, police records, education data, and healthcare data. Platforms that are able to access, aggregate, and curate such data sets do not exist; this is an enormous barrier to progress. Tools that build on such data sets – for example, basic methods for federation, inferences, and so forth can only be meaningfully developed once such infrastructure tools are available.

**Longitudinal Perspective:** Many of the problems related to public welfare are not "solvable" in the sense that they have a clear end point where they cease to be a problem. Further, the effect of a particular innovation or intervention may only become apparent over a period of years or decades and may be difficult to prove, as a controlled trial may be difficult or impossible to construct. Projects need to have a long-term structure, with appropriate intermediate goals, to avoid short-term fixes, or quick, but ephemeral, "feel-good" stories.

**Legal, Regulatory, Compliance:** No list of barriers would be complete without acknowledging that there are many legal and regulatory hurdles for many of these projects. Access to data, and to populations to evaluate against will require substantial investment of time, planning, and resources to have an effect. Creating frameworks for ethical evaluation of costs and benefits must be established. Understanding the impact of innovations will require an understanding of the level of compliance, and possibly methods to manage or pivot solutions in response to perception, trust, and compliance of the target population.

## Fundamental Issues

Although AI for public welfare draws on many common themes and ideas of AI research more broadly, there are ARTIFICIAL INTELLIGENCE FOR SOCIAL GOODalso some key differences. First and foremost, public welfare is ultimately about information and decisions that directly affect the lives of individuals. As a result, privacy issues, transparency and traceability of data collection and decision-making, and understanding of social context are factors that must be considered within the research context. Issues surrounding data bias and uncertainty have direct implications to fairness and the evaluation of the utility of possible decision paths. Indeed, formalizing impact of decision on the individual, and understanding how individual preferences play into those decisions is as important as the inferential framework that leads to those decisions.

A second significant challenge is to understand and frame the problem with respect to the sociodynamics of the population in question, and the organizational constraints that shape possible responses. For example, broad-based surveillance of infant health may identify areas where lead poisoning is a substantial health risk, but the ability of a government to respond – by remediation, by education, by identifying and supporting relocation of at-risk populations – will vary by locality and by the willingness of the population involved.

With these challenges in mind, there are numerous basic and applied research problems to be solved. Some of these have been raised above. Others include: 1) data analytics and machine learning models that are robust to systematic bias, missing data, and data heterogeneity; 2) the development of models or simulations that are sufficiently predictive to inform decision-making, and which also can then be adapted "closed-loop" as additional data is collected with time; 3) advanced models of decision-making and planning that incorporate social dynamics, resource constraints, and utility models for multiple actors; 4) consistent, cost-effective, and scalable models for measurement or data collection; and 5) methods for causal reasoning and explanation.

It is important to emphasize that these core AI problems have to be married with a set of broader computer-science innovations. For example, much of the data in question will be personal and sensitive; scalable progress will depend on privacy preserving methods, particularly those that are robust to federation of data from multiple disparate sources. Additionally, user-centered design approaches taking into account the individuals or populations will be essential for adoption.

## Opportunities

There are innumerable opportunities to advance work in AI for public welfare, for example:
- Better data collection, digitization, and curation, particularly around urgent priorities.
- Better federation and integration of data sources currently not being used together.
- Better models and predictions of individual behaviors to support existing interventions
- Better evaluation of existing and historical policies to understand their implications vis-a-vis enablement of AI advances.

## More advanced AI capabilities in these areas could contribute to public welfare in many ways, including:

**Public safety:** Better data on the location and activities of first responders would quickly (over the space of months) create a dataset that could be mined to create predictive algorithms to better deploy first responders.

**Transportation:** Using individual public transit and other transportation data (uber, bikeshare, etc.) would allow researchers to better understand mobility patterns of people, to understand gaps in transit with respect to citizenry needs and also to assess the impact of policy changes through data-enabled simulation analysis.

**Public Health:** School records, employment records, health records, and neighborhood level population statistics could be combined to build better predictive models and better detection of high-risk health events – e.g. women who may be at risk of adverse birth events to target human services programs and resources, men who are under stress due to long-term health and employment issues where intervention may be warranted, and so forth.

**14**

**Public Welfare:** Public data feeds and public reporting, combined with the growing body of surveillance data would potentially identify individuals or populations that are in danger of becoming homeless, incarcerated, or otherwise taking a path likely to lead to both personal and social costs. This could also be a fertile ground to study the interaction of prediction models and the effectiveness of interventions - i.e. heterogenous treatment effects, similar to the phenomena of behavior change due to clinical decision support.

**Education:** Social feeds, school records, and well constructed social network-informed behavioral models could be used to detect school populations that are at risk of falling below grade level and thus where to deploy interventions that could influence and change behavior much earlier than is now currently possible.

## Research Challenges

**Humans In The Loop:** All of the ideas above have an inherent complexity that goes well beyond what can be inferred from data sets – for example, social behavior is shaped by government policies, by "what's hot," by news events, by the weather, or by many other phenomena that may be idiosyncratic to a particular locale, subpopulation, or point in time. Thus, all of the systems above must really be considered as "human-in-the loop" systems that will almost always involve human policy and decision-makers, and of course operate on a specific population. Thus, research is needed in understanding how to develop these systems to be maximally effective and enabling within the context of system or organization.

**Measuring Engagement:** Much of what we'd like to affect cannot be directly measured or controlled. AI research will need to move beyond prediction of discrete (easily measurable) outcomes and instead use models/algorithms to optimize hidden variables such as engagement/happiness for students in school, people in a community, or in government. To do so, we will need to create a set of concepts, methods, and tools and methods that AI, Social Science, and policy specialists agree on and which form a "franca lingua" for work in this area.

## A Grand Challenge

A gross generalization of all of the above themes is that public good is a complex system of systems – information about welfare impacts education impacts law enforcement impacts health, and so forth. Just the implementation of a healthcare system or a transformation system involves the integration of many interacting components. We need to create methods to abstract even further and develop methods to integrate multiple AI systems, which collectively monitor, detect, diagnose, and adapt within their own specific domains.

From this, we can frame the ultimate grand challenge of AI for public welfare: Can we create tools that automatically and proactively identify problem causes, propose policy solutions, and predict consequences of those (potentially cross-issue) policies? Can we identify solutions that a person may not immediately come up with because of the ability of a system of systems model to look across domains and see linkages that no single individual could?

## Cross Cutting Issues

This report started out by asserting that AI can be a major force for social good; but that to make it such a force, we need to shape this new technology and the questions we use to inspire young researchers. In this report, the term "social good" is intended to focus AI research on areas of endeavor that are to benefit a broad population in a way that may not have direct economic impact or return, but which will enhance the quality of life of a population of individuals through education, safety, health, living environment, and so forth. In general these are areas of work that have not benefited from AI research, but are nonetheless important for societal benefit.

At the end of this report, we now come to some key observations:

◗ *AI for Social Good*: We first observe through the research reported in this report that there is ongoing work, be it in urban computing, sustainability, health, public welfare, leading the way for applying AI for Social Good. AI research is already being shaped in applications for social good; but this trend needs to be encouraged.

- *Use inspired research:* Research led from applications, from actual use, is important for this area of work, and in shaping AI for Social Good. Use inspired work in this area will lead to questions that are crucial to making a social impact, while also providing innovative research possibilities. For example, consider working with social networks in low resource communities; given unreliable and uncertain access to smart phones or other technologies, research questions that arise may be fundamentally different, often arising out of insufficient data about the networks. This is but one example, but the essential point is to drive the research from its use.

- *Interdisciplinary teams and new styles of research:* Research on AI for Social Good necessarily will require research with interdisciplinary teams, where part of the team is rooted firmly in the domain discipline.

  A novel aspect of this interdisciplinary work is new methods for evaluating interdisciplinary work and measuring impact. It is not sufficient then to only claim an AI contribution by showing an improvement in a modeling technique's efficiency or testing in simulations; we would need to measure real impact in the field in terms of what was truly accomplished. Such measurements require time and effort, and it is not typically ready on a six month AI research conference cycle. Nonetheless, it is important that this type of science be encouraged and allowed to thrive (via new publication venues, prestigious awards, etc.). Traditionally AI publication venues focus on methodological advances, and impact driven work is relegated to "application conferences". This would need to shift if in order to allow young scientists to get involved in research with social impact, to allow the scientists to build up careers in this interdisciplinary research space.

- *Interpretability, transparency, accountability are important:* With interdisciplinary research and impact on society comes the burden that the resulting AI models be interpretable and transparent. Not only may users and collaborators not accept results that are output from a "black box" un-interpretable algorithm, but there is also a real danger that the black box may be using flawed or even illegal means of arriving at its conclusions. For example, a "black box" AI algorithm may rely on racist or sexist inferences for its conclusions (arriving there due to biased input). Thus interpretability and transparency of the algorithms will remain key requirements as we go into the future with AI for Social Good.

- *Human-AI boundary:* As we apply AI for Social Good applications, many of these applications are seen to be decision aids, assisting the human. Because these applications might be in domains with vulnerable populations (but even if not so), the right human-AI interface is important to consider. The issue is not just the HCI aspects, but in a fundamental sense where to draw the boundary between AI and the human interacting with the AI.

  Key principles of such human-AI interaction are an important issue for future work. Some principles are becoming apparent. For example, one such principle is respect for people's autonomy. There are situations where humans have superior control, insight knowledge; for the AI system to have a human as a subordinate following its commands will not make sense. In such situations, instead, the human must be in charge. There are of course many others that veer into broader issues of safety in control, and ensuring justice (in the sense of making the AI's benefit available to all segments of society). These will remain important issues to consider into the future.

In summary, like many technologies before it, AI is a family of tools that will find their way into a broad spectrum of applications, many of which we cannot today imagine. However, as the discussion above demonstrates, with appropriate forethought and incentives, AI can become a tool that enhances our quality of life at a personal, national, and global level. In this regard, we hope this report inspires new ideas and approaches that leverage support for basic AI research with opportunities to work with state and local governments and other nonprofit organizations.

Finally, we note that many of the problems and solutions are uniquely local – many at the level of a city or community. Thus, AI for Social Good also provides a unique opportunity for technology researchers to personally engage with their local communities and, by doing so, concretely educate the public about the technology, its limitations, and its potential benefits. We hope this report will succeed in inspiring the AI research community to seek out and capitalize on these opportunities.

## Artificial Intelligence (AI) for Social Good Workshop

There has been an increasing interest in Artificial Intelligence (AI) in recent years. AI has been successfully applied to societal challenge problems and it has a great potential to provide tremendous social good in the future. At this workshop, we discussed the successful deployments and the potential use of AI in various topics that are essential for social good. There were over 300 participants at the workshop, including 87 females and 75 individuals from industry, with an additional 3,500 viewers on the livestream.

Workshop website- http://cra.org/ccc/events/ai-social-good/

Videos from the workshop- http://cra.org/ccc/artificial-intelligence-social-good-speakers/

Slides from each presentation- http://cra.org/ccc/events/ai-social-good/#agenda

**Post-workshop AI Roundtable Discussion**

Roy Austin, White House
Guru Banavar, IBM
Tanya Berger-Wolf, University of Illinois at Chicago, IBEIS.org
Randal Bryant, Carnegie Mellon University
Evan Cooke, OSTP
Thomas Dietterich, Oregon State University/AAAI
Ann Drobnis, Computing Community Consortium
Eric Elster, Uniformed Services University of the Health Sciences & the Walter Reed National Military Medical Center
Fei Fang, University of Southern California
Ed Felten, OSTP
Ned Finkle, NVIDIA
Melissa Flagg, DoD
Michael Garris, NIST
Rayid Ghani, University of Chicago
Carla Gomes, Cornell University, Institute for Computational Sustainability
Amy Greenwald, Brown University
Greg Hager, Johns Hopkins University
Verity Harding, Google
Peter Harsha, Computing Research Association
Dan Hoffman, Chief Innovation Officer
Vasant Honavar, Pennsylvania State University
Eric Horvitz, Microsoft Coporation
Henry Kautz, University of Rochester
Michael Littman, Brown University
Terah Lyons, OSTP
Alexander Macgillivray, OSTP
Mary Miller, Department of Defense
Elizabeth Mynatt, Georgia Tech / CCC
Jennifer Neville, Purdue University
Dale Ormond, DoD
Lynn Overmann, OSTP
Lynne Parker, NSF
David Parkes, Harvard University
Suchi Saria, Johns Hopkins University
Reuben Sarkar, DOE
William Scherlis, Carnegie Mellon University
Jason Schultz, OSTP
Stephen Smith, Carnegie Mellon University
Suhas Subramanyam, OSTP
Milind Tambe, University of Southern California
Pascal Van Hentenryck, University of Michigan
Hanna Wallach, Microsoft Research

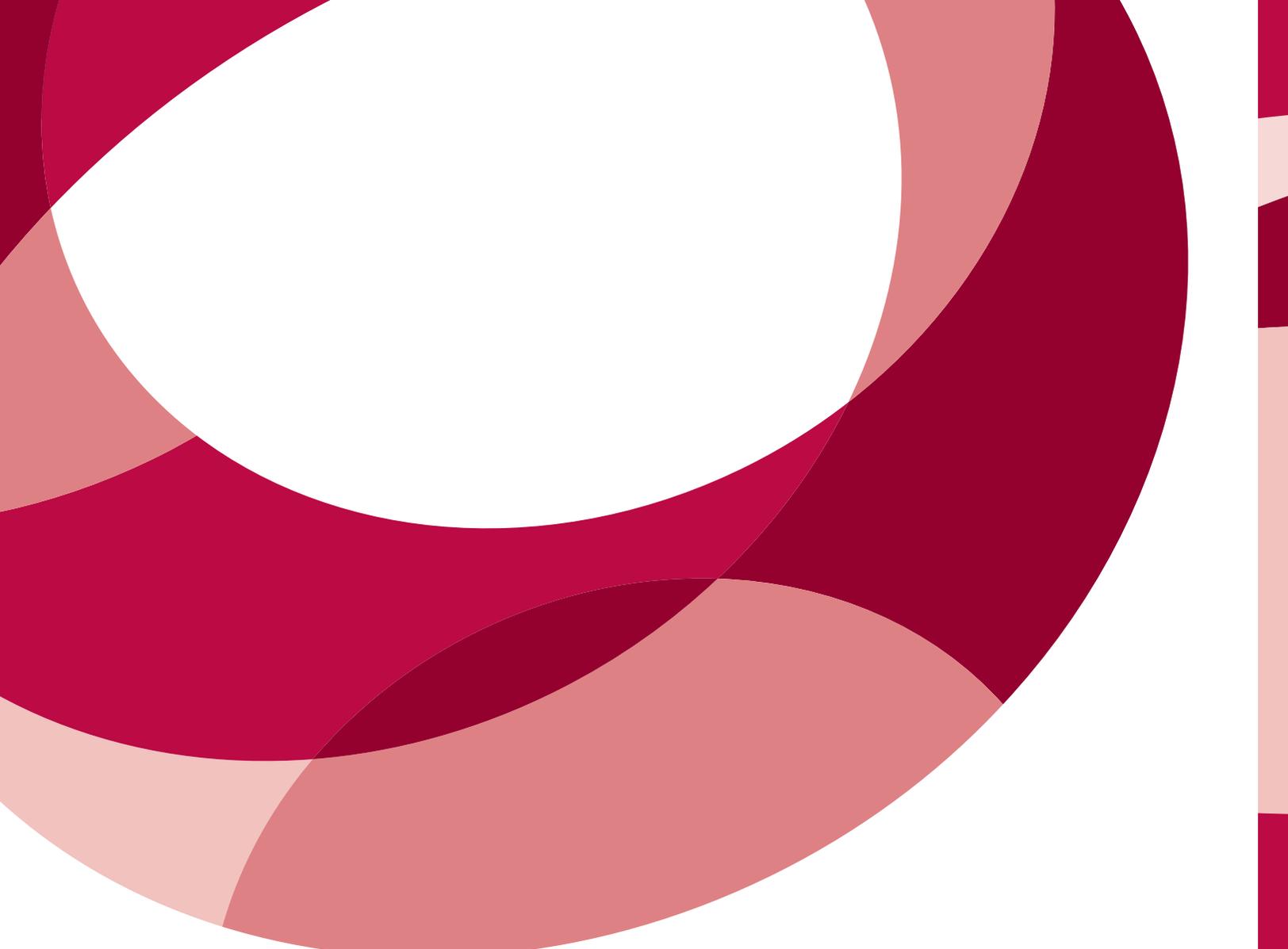

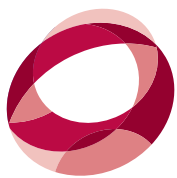

**CCC**
Computing Community Consortium
Catalyst

1828 L Street, NW, Suite 800
Washington, DC 20036
P: 202 234 2111 F: 202 667 1066
www.cra.org cccinfo@cra.org